\documentclass{cernrep}
\usepackage{texnames}
\usepackage[T1]{fontenc}
\usepackage{color}
\newcommand{\cc}{\,\mathrm{cm^{-3}}}
\newcommand{\wcm}{\,\mathrm{W\,cm}^{-2}}
\newcommand{\mic}{\,\mu\mathrm{m}}

\newcounter{query}

\renewcommand{\rmd}{{\rm d}}

\begin{document}
\title{Plasma Injection Schemes for Laser--Plasma Accelerators}

\author{J. Faure}

\institute{Laboratoire d'Optique Appliqu\'ee, ENSTA-CNRS \'Ecole Polytechnique, Palaiseau, France}

\maketitle 

\begin{abstract}
Plasma injection schemes are crucial for producing high-quality electron beams in laser--plasma accelerators. This article introduces the general concepts of plasma injection. First, a Hamiltonian model for particle trapping and acceleration in plasma waves is introduced; ionization injection and colliding-pulse injection are described in the framework of this Hamiltonian model. We then proceed to consider injection in plasma density gradients.
\end{abstract}

\section{Introduction}
In accelerator physics, the starting point of a high-energy machine is always the injector. The injector can be extremely important; it usually delivers a particle beam at the$ \UMeV$  level, and its characteristics---such as emittance, bunch duration and energy spread---strongly impact the parameters of the final beam. By analogy, it seems only natural that laser--plasma accelerators should also include an injection stage; this would permit (i)~decoupling of the injection and acceleration stages, (ii)~more control over the beam parameters and the possibility to tune them independently, and (iii)~better stability of the beam. However, injecting electrons into the plasma wakefield accelerating structure is not an easy task: the wavelength of the accelerating structure is on the order of $\lambda_{\rm p}\simeq10\:\mu$m; here, the plasma wavelength is $\lambda_{\rm p}=2\pi c/\omega_{\rm p}$ (where $c$ is the speed of light and $\omega_{\rm p}=(n_{\rm e} e^2/\epsilon_0m_{\rm e})^{1/2}$ is the plasma frequency), so that $\lambda_{\rm p}=10$--$30\:\mu$m for electron plasma densities in the $10^{18}$--$10^{19}\:$cm$^{-3} $ range. The production of monoenergetic electron bunches in such a micrometre structure requires that the injected beam have a duration shorter than $\lambda_{\rm p}/c=30$--$100\:$fs and that it be focused to a few microns. In addition to these stringent requirements, the injected particle beam should be synchronized at the femtosecond level with the laser pulse driving the wakefield, which also poses a considerable experimental challenge. The production of such short and well-synchronized bunches is at the edge of conventional radio-frequency (r.f.) accelerator technology, and experiments using external injectors have not yet succeeded in providing high-quality electron beams with narrow energy spreads \cite{ever94,amir98}.

In many laser wakefield acceleration experiments, electrons are \textit{self-injected} into the accelerating structure: when the plasma wave amplitude reaches very high levels (close to the wave-breaking threshold), background plasma electrons can be injected into the plasma waves \cite{mode95,umst96,faur04,gedd04,mang04}. This is somewhat analogous to dark current in a r.f.\ accelerator: at high field strengths, the cavities start to release electrons. This self-injection mechanism can lead to the production of high charge and, on occasion, narrow energy spreads in the range of a few percent to 10\% can be achieved. However, it has proven difficult to obtain high-quality, i.e.\ narrow, energy spreads in a stable manner using this method. In addition, electron injection results from a succession of nonlinear effects, such as relativistic self-focusing \cite{sun87,bori92} of the laser pulse, spectral broadening and self-steepening \cite{esar00,gord03,faur05,schr10}. This is the reason that self-injection is rather difficult to control precisely and does not allow tuning of the injected beam parameters. In this context, many researchers in the field of laser--plasma accelerators have started to develop injection methods in which the injected electrons originate from the plasma itself (as opposed to external injection requiring an existing electron source).

While several methods have been proposed for injecting plasma electrons into an existing plasma wakefield, they can be summarized by the following general principles.
\begin{itemize}
\item Create electrons at the right phase in the wakefield. Even if an electron is created at rest, it can end up being trapped provided that it is born at the appropriate phase. This is the idea behind ionization injection.

\item Give an initial kick to electrons so that their initial longitudinal velocity is high enough for trapping. This is analogous to surfers paddling in order to gain momentum and catch the wave. Several methods are based on this idea, including injection by colliding pulses.

\item Slow down the wakefield to facilitate trapping. Decreasing the phase velocity of the wakefield locally can be achieved by tailoring the plasma density, for example, and it leads to a controlled injection mechanism. Injection in downward density ramps is based on this idea.
\end{itemize}

In this article we will review some of these injection mechanisms. In Section \ref{secHamiltonian}, a Hamiltonian formalism based on \Refs~\cite{esar95,esar96} is derived in order to find the trapping threshold for an electron interacting with an intense laser pulse and a plasma wave. In Section~\ref{secIoniz}, the findings from the Hamiltonian model will be applied to the particular case of ionization injection where high-$Z$ atoms provide the source of injected electrons. In Section~\ref{secCollid} we show how electrons can be injected using colliding and counter-propagating laser pulses. Finally, in Section~\ref{secDensity}, we show how electrons can also be trapped in wakefields with decreasing phase velocities.

\section{One-dimensional Hamiltonian model}\label{secHamiltonian}

\subsection{Assumptions of the model}
The goal of this section is to derive
analytically the trajectories of electrons in a laser field and a plasma wave \cite{esar96}. In
particular, some electrons can be trapped and accelerated in the
plasma wakefield. The theoretical framework is the following.

\begin{figure}[htbp]
\begin{center}
\includegraphics[width=8cm]{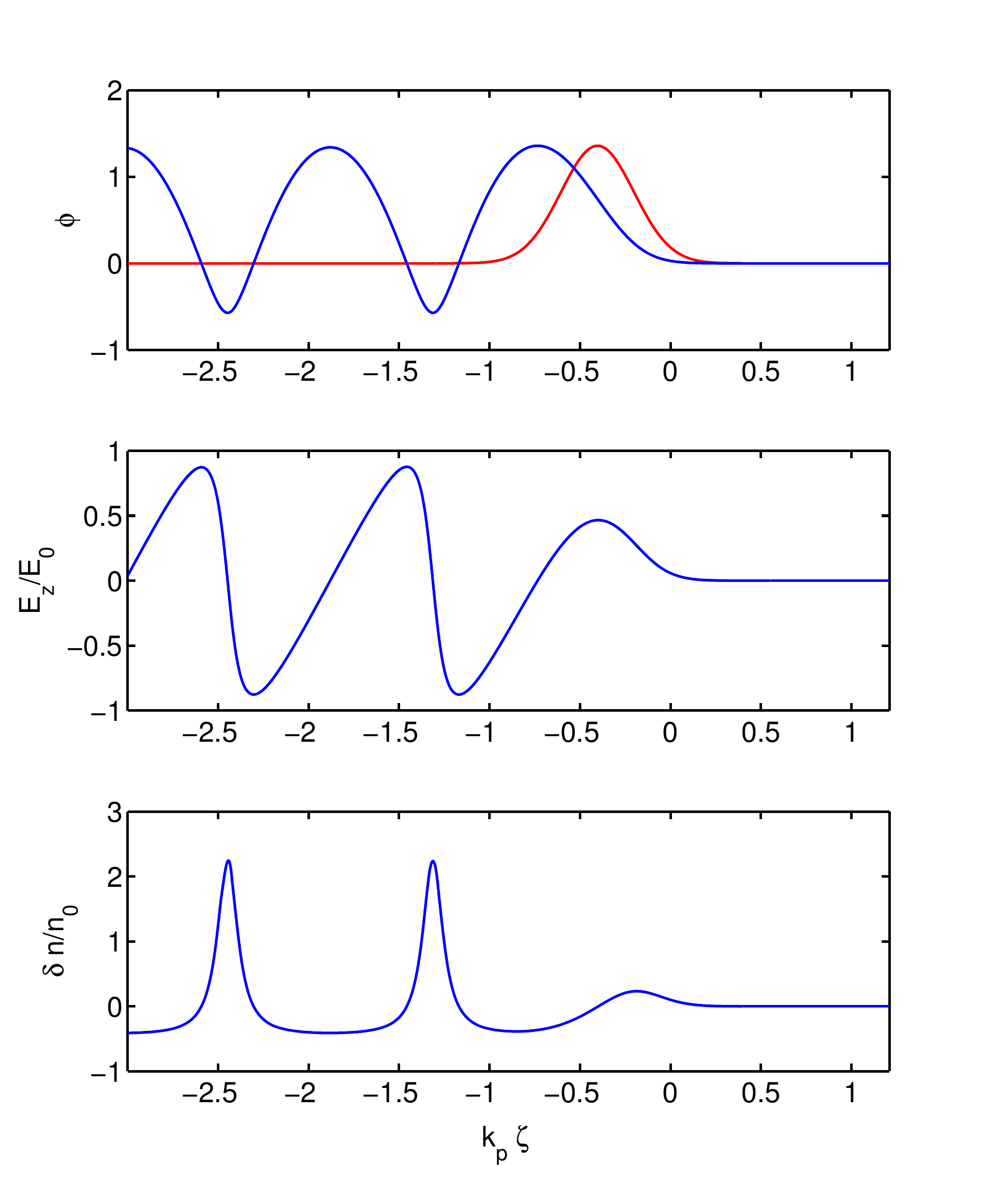}
\caption{The top panel plots the laser amplitude $a$ (red) and the wakefield potential $\phi$ (blue); the middle panel shows the corresponding normalized electric field $E_z/E_0$ (where $E_0=m_{\rm e}c\,\omega_{\rm p}/e$ is the wave-breaking field). The bottom panel shows the electron density perturbation $n(\zeta)/n_0-1$.}
\label{figwakefield}
\end{center}
\end{figure}

\begin{itemize}
    \item We start with a 1D model: we consider only the motion of electrons along the longitudinal coordinate $z$, and neglect the role of the radial electric fields. In
    this case, the wakefield potential $\phi$ depends only on $z$ and $t$.
    \item For simplicity, we assume that the laser pulse does not
    change during its propagation. Consequently,
    the plasma wakefield is also stationary along the propagation.
    This is important because it will allow us to use a
    conservation-of-energy law.
     \item The plasma is modelled by an electron fluid. This fluid is
    described by macroscopic quantities such as its density $n(\mathbf{r},t)$ and
    velocity $\mathbf{v}(\mathbf{r},t)$. Let us note that in
    such a model, kinetic effects (e.g.\ trapping, wave-breaking) are not
    taken into account.
     \item Plasma ions are assumed to be immobile. This is
    justified when the typical time for ion motion ($\omega_{\rm pi}^{-1}$) is large
    compared to the driver pulse duration
    (i.e.\ $\tau\ll\omega_{\rm pi}^{-1}$).
        \item The electron fluid is cold. In the case of a laser driver, this is justified
    when the quiver velocity of electrons in the laser field is orders of magnitude greater
    than the thermal velocity: $v_{\rm osc}\simeq~eE_{\rm laser}/(m_{\rm e}\omega_0)\gg
    v_{\rm th}=(k_{\rm B} T_{\rm e}/m_{\rm e})^{1/2}$.
\end{itemize}
The laser pulse is represented by its normalized vector potential
$\mathbf{a}=e\mathbf{A}/m_{\rm e}c$, where $\mathbf{A}$ is the laser vector potential. 
For a pulse propagating along the $z$-axis and polarized
along the $x$-axis,
\begin{equation}
\mathbf{a}=\hat{a}(z,t)\cos(k_0z-\omega_0t)\,\mathbf{e}_x,
\end{equation}
where $k_0$ and $\omega_0$ are the wavevector and frequency of
the laser electromagnetic field, and $\hat{a}$ is an envelope
function that represents the longitudinal shape of the pulse.
We assume a Gaussian shape for $\hat{a}$
so that $\hat{a}^2(\zeta)=a_0^2 \exp(-\zeta^2/L_0^2)$, where $\zeta=z-v_{\rm g}t$ with $v_{\rm g}$ being the laser group velocity, $L_0$ is the laser pulse length, and $a_0$
is the laser peak amplitude, $a_0=8.6\times10^{-10}\sqrt{I\,[\wcm]\,\lambda^2\,[\mathrm{\mu m}]}$, with $I$ and $\lambda$ being the laser intensity and the central wavelength, respectively. The plasma wakefield is represented by its normalized potential $\phi=e\Phi/m_{\rm e}c^2$, which is obtained by solving the equation
\begin{equation}\label{eqphi}
\frac{\partial^2\phi}{\partial
\zeta^2}=k_{\rm p}^2\gamma_{\rm p}^2\left[\beta_{\rm p}\Bigg(1-
\frac{1+a^2}{\gamma_{\rm p}^2(1+\phi)^2}\Bigg)^{-1/2} -1\right].
\end{equation}
Here, $\gamma_{\rm p}$ is the Lorentz factor corresponding to the plasma wave phase velocity $v_{\rm p}$: $\gamma_{\rm p}=(1-v_{\rm p}^2/c^2)^{-1/2}$. It is also assumed that the plasma wave velocity is given by the group velocity of the excitation laser: $v_{\rm p}\simeq v_{\rm g}$ and $\gamma_{\rm p}\simeq \gamma_{\rm g}$. Finally, the plasma wavevector is $k_{\rm p}=\omega_{\rm p}/v_{\rm p}$. \Eref{eqphi} is the nonlinear plasma wave equation, which is valid even for $a>1$, i.e.\ for relativistic laser intensities with $I>10^{18}\:\wcm$. Figure~\ref{figwakefield} shows the result for a 20~fs laser pulse with $a_0=2$ and a plasma density of $n_{\rm e}=7\times~10^{18}\:\cc$. Note that the laser field was averaged over the fast time-scale at $\omega_0$ for clarity (keeping the fast frequency does not affect our conclusions). Notice the characteristic nonlinear shape of the plasma density perturbation, which in turn causes a nonlinear longitudinal electric field with a sawtooth shape.

\subsection{Hamiltonian for an electron interacting with a laser and a plasma wave}

Let us consider an electron with Lorentz factor $\gamma$ in a 1D wakefield, represented by the normalized potential $\phi$. Its Hamiltonian reads $H =\gamma-\phi(z-v_{\rm g}t)=\sqrt{1+u_\bot^2+u_z^2}-\phi(z-v_{\rm g}t)$, where $u_{\bot}=p_{\bot}/(m_{\rm e}c)$ and $u_{z}=p_{z}/(m_{\rm e}c)$ are the transverse and longitudinal normalized linear momenta, respectively. The Hamiltonian depends on $z$ and $t$ in a particular manner: it depends only on the variable $\zeta=z-v_{\rm g}t$. We can use this
fact to change variables using a canonical transformation
$(z,u_z)\rightarrow (\zeta,u_z)$. We use a second-type generating
function $F_2(z,u_z)=u_z(z-v_{\rm g}t)$; the transformation satisfies $\zeta=\partial F_2/\partial
u_z=z-v_{\rm g}t$ and $u_z=\partial F_2/\partial z$, and the new Hamiltonian
is given by $H'=H+\frac{1}{c}\frac{\partial F_2}{\partial t}$. In this case, the new Hamiltonian (denoted simply by $H$ from now on) is
\begin{equation}\label{EqHbasic}
    H=\sqrt{1+u_\bot^2+u_z^2}-\phi(\zeta)-\beta_{\rm p}u_z
\end{equation}
where $\beta_{\rm p}=v_{\rm p}/c$.

This Hamiltonian has several constants of motion. We now introduce the canonical momentum $\mathbf{P}=\mathbf{p}+q\mathbf{A}$. In normalized units, the canonical momentum is denoted by $\mathbf{U}$ and we see that $U_z=u_z$ and $U_\bot=u_\bot-a$ ($q=-e$ for electrons), because in 1D the laser field has only a transverse component. Writing the Hamiltonian with just the transverse canonical momentum gives
\[
H =\sqrt{1+(U_\bot+a)^2+u_z^2} -\phi(\zeta)-\beta_{\rm p} u_z.
\]
From Hamilton's equations, one finds that in 1D, the transverse canonical momentum is conserved:
\begin{equation*}
\dot{U}_\bot=-\frac{\partial H}{\partial r_\bot}=0 \,\implies \, u_\bot(\zeta)-a(\zeta)=\mathrm{const}.
\end{equation*}

\begin{figure}[h]
\begin{center}
\includegraphics[width=8cm]{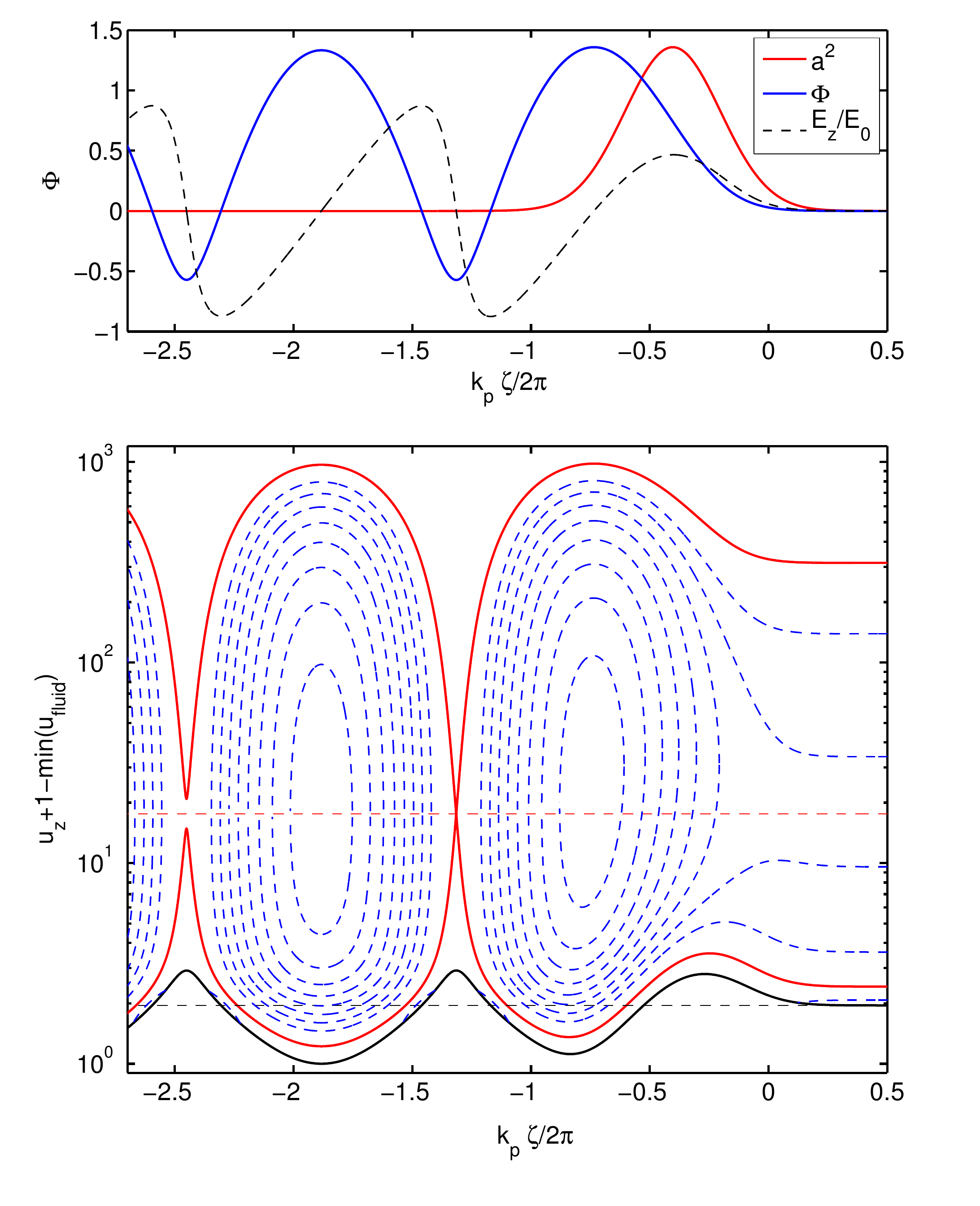}
\caption{The top panel plots the laser amplitude $a^2$ (red), the wakefield potential (blue), and the longitudinal electric field (black dashed); the bottom panel shows the associated electron trajectories in phase space. The black line in the bottom panel represents the fluid orbit, the red line the separatrix, and the dashed blue lines various trapped orbits. The $y$-axis is in log-scale and, to avoid negative values, we have plotted $u_z+1-\min(u_{\rm fluid})$, where $u_{\rm fluid}(\zeta)$ is the fluid orbit. The horizontal black dashed line represents $u_z=0$, while the horizontal red dashed line represents $u_z=\beta_{\rm p}\gamma_{\rm p}$. The parameters are $a_0=2$, $n_{\rm e}/n_{\rm c}=0.44\%$, $\lambda=0.8\:\mathrm{\mu m}$ and $\tau=20\:\mathrm{fs}$.}
\label{figsep}
\end{center}
\end{figure}

In the case of an electron initially at rest in front of the laser pulse, $\zeta_{\rm i}=+\infty$, this gives the important result that $u_\bot(\zeta)=a(\zeta)$. The other constant of motion is simply the energy: this Hamiltonian does not depend on time but only on
$\zeta$; as a consequence, the energy of the system is conserved and
$H$ is conserved along an electron trajectory. Hence, for an electron with initial energy $H_0$, one can solve for its longitudinal momentum $u_z$ to get
\begin{equation}\label{EqTraj}
u_z=\beta_{\rm p}\gamma_{\rm p}^2(H_0+\phi)\pm\gamma_{\rm p}\sqrt{\gamma_{\rm p}^2(H_0+\phi)^2-\gamma_\bot^2}
\end{equation}
with $\gamma_\bot^2=1+u_\bot^2$. Once $a(\zeta)$ and $\phi(\zeta)$ are known, this equation gives the electron trajectory in $(\zeta,u_z)$ phase space. Figure~\ref{figsep} shows various electron trajectories that were obtained for different initial conditions.
For electrons initially at rest in front of the laser pulse, i.e.\ $\zeta_{\rm i}=+\infty$ and $u_z(\zeta_{\rm i})=u_\bot(\zeta_{\rm i})=~0$, the Hamiltonian is $H_0=1$. The trajectory of such electrons is referred to as the fluid orbit and corresponds to the trajectory of plasma background electrons that contribute to the formation of the plasma wakefield (black line in lower panel of \Fref{figsep}). These electrons are not trapped and oscillate in the plasma wakefield with low energies.
The separatrix is the special trajectory which separates the trapped from the untrapped orbits. It can be found by considering an electron moving along $z$ with $v_z=v_{\rm p}$ (or $u_z(\zeta_{\min})=\beta_{\rm p}\gamma_{\rm p}$) and located initially at a minimum of the potential $\phi(\zeta_{\min})=\phi_{\min}<0$, i.e.\ at a node of the electric field. Conservation of canonical momentum reads $u_\bot(\zeta_{\min})=a(\zeta_{\min})$, so that the Hamiltonian on the separatrix is
\begin{equation}
H_{\rm sep}=\frac{\sqrt{1+a^2(\zeta_{\min})}}{\gamma_{\rm p}}-\phi_{\min}.
\end{equation}
This trajectory is represented by a red line in the lower panel of \Fref{figsep}. The dashed blue lines represent some typical trapped orbits for which electrons gain significant energy in the plasma wave. Trapped orbits are obtained when the Hamiltonian satisfies $H_{\rm i}\leqslant H_{\rm sep}$. To prove this, let us consider a trapped electron behind the laser pulse at phase $\zeta_{\rm i}$ and moving with  velocity $v_z(\zeta_{\rm i})=v_{\rm p}$. At this point in phase space, the plasma wave potential is such that $\phi(\zeta_{\rm i})=\phi_{\rm i}\geqslant\phi_{\min}$. In addition, following \Eref{EqHbasic}, conservation of the Hamiltonian can be written  as $H_{\rm i}=(1+\beta_{\rm p}^2\gamma_{\rm p}^2)^{1/2}-\phi_{\rm i}-\beta_{\rm p}^2\gamma_{\rm p}=1/\gamma_{\rm p}-\phi_{\rm i}$. Since
$\phi_{\rm i}\geqslant\phi_{\min}$, we get $H_{\rm i}\leqslant H_{\rm sep}$, which is a necessary and sufficient condition for trapping.

\subsection{Trapping thresholds}
\begin{figure}[h]
\begin{center}
\includegraphics[width=14cm]{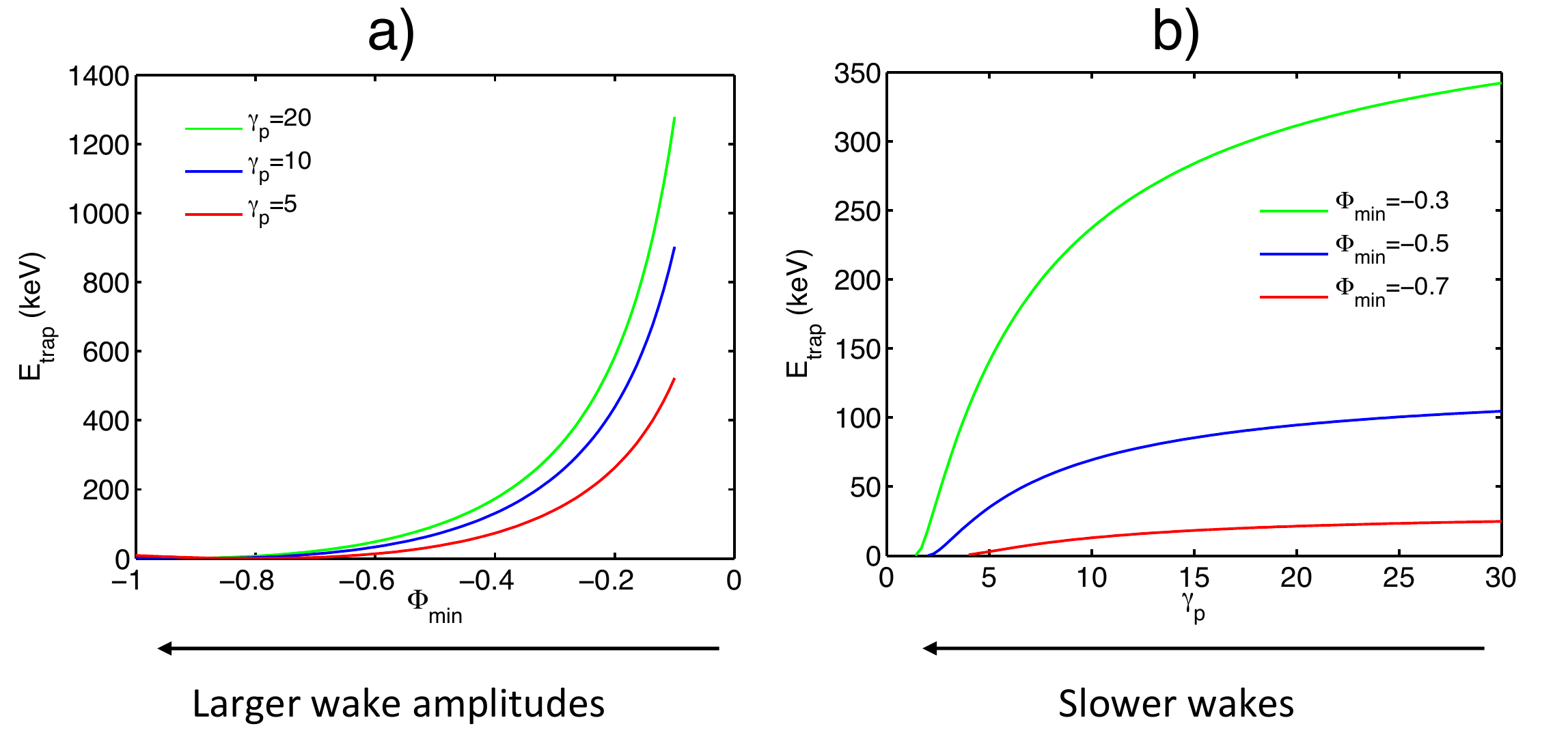}
\caption{(a) Trapping thresholds plotted as a function of wake amplitude (where $\phi_{\min}$ represents the minimum of the wake potential), for three different wake phase velocities. (b)~Trapping thresholds plotted as a function of $\gamma_{\rm p}$, the Lorentz factor associated with the wakefield velocity, for three different wake amplitudes.}
\label{figtrap}
\end{center}
\end{figure}

Now that the basic trajectories have been established, it is possible to determine the trapping threshold of plasma electrons in a plasma wave. To do so, we consider that an electron located in front of the laser pulse will be trapped in the wakefield if its longitudinal momentum satisfies $u_z(+\infty)>u_{z}^{\rm sep}(+\infty)$, which simply means that the initial momentum has to be larger than the initial momentum on the separatrix, $u_{z}^{\rm sep}(+\infty)$. According to \Eref{EqTraj}, this is simply $u_{z}^{\rm sep}(+\infty)=\beta_{\rm p}\gamma_{\rm p}^2H_{\rm sep}-\gamma_{\rm p}\sqrt{\gamma_{\rm p}^2H_{\rm sep}^2-1}$. Therefore, an electron with initial energy $E>E_{\rm trap}$ will be trapped and accelerated in the wakefield, where
\begin{equation}
E_{\rm trap}=m_{\rm e}c^2\Bigl(\sqrt{1+\{u_z^{\rm sep}(+\infty)\}^2}-1\Bigr).
\end{equation}
Figure \ref{figtrap} shows the variation of the trapping threshold with the plasma wave amplitude $\phi_{\min}$, as well as its variation with the plasma wave Lorentz factor $\gamma_{\rm p}$. Clearly, trapping is easier for high-amplitude plasma waves and/or for small phase velocities (small values of $\gamma_{\rm p}$). Note that as $\phi_{\min}\rightarrow -1$, the trapping threshold $E_{\rm trap}$ tends toward zero. This is the onset of wave-breaking: the plasma wave amplitude becomes so high that all plasma electrons that were initially at rest get injected into the plasma wave. As $\phi_{\min}\rightarrow -1$, the longitudinal electric field reaches the cold wave-breaking limit, $E_{\rm WB}=\sqrt{2}(\gamma_{\rm p}-~1)^{1/2}E_0$, where $E_0=m_{\rm e}c\,\omega_{\rm p}/e$ was defined earlier. Thus, this model is able to give a 1D picture of self-injection, although one has to keep in mind that in experiments self-injection is a 3D process. More complicated models are necessary to fully capture the physics of self-injection in 3D \cite{kost04}.

\section{Ionization injection}\label{secIoniz}
\begin{figure}[h!]
\begin{center}
\includegraphics[width=16cm]{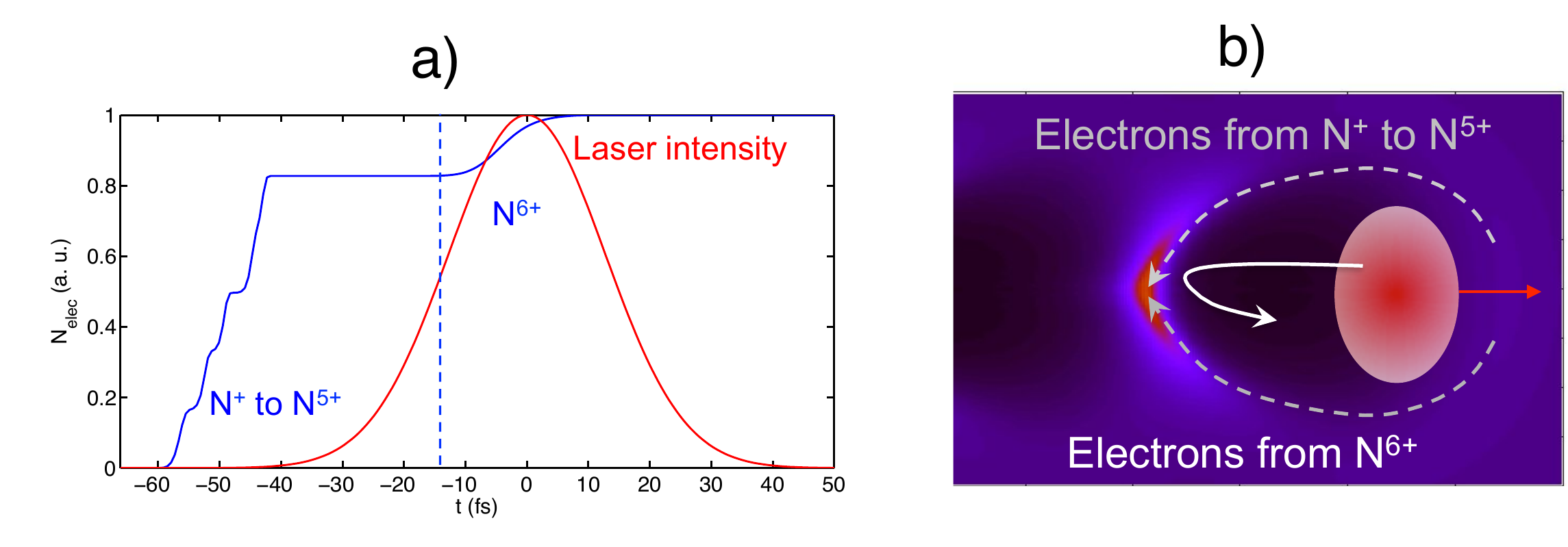}
\caption{Principle of ionization injection. (a)~Ionization of nitrogen by an intense laser pulse at $10^{19}\:\wcm$; the first five electrons, from N$^+$ to N$^{5+}$, appear at the front of the laser pulse, while electrons from N$^{6+}$ appear in the middle of the laser pulse. (b)~Schematic showing the different classes of trajectories for electrons born at the front of the pulse (fluid trajectories) or in the middle of the pulse (trapped trajectories).}
\label{figioniz}
\end{center}
\end{figure}

As we have just seen, unless the wakefield reaches wave-breaking amplitudes, it is not possible for plasma electrons to be injected unless a different method is used. Ionization injection is probably the easiest method for injecting electrons in a wakefield \cite{pak10,mcgu10}. The idea is to use a high-$Z$ gas so that the first levels of ionization occur at low intensity (typically below $10^{16}\:\wcm$). These electrons are born at rest in a region where the laser intensity is relatively low and provide the electrons which form the plasma wave and follow fluid orbits. On the other hand, ionization from inner shells occurs at higher intensities (typically for $I>10^{18}\:\wcm$), so these electrons are born at rest in regions of strong fields. It follows that they are born inside the plasma wave, at a totally different phase from fluid electrons. The whole challenge is to put these electrons on trapped orbits so that they will be injected and accelerated. Figure~\ref{figioniz} illustrates the different classes of trajectories of  electrons born at the front of the pulse and in the middle of the pulse. In this section, we follow the approach of \Refs~\cite{chen10,pak10}.

\begin{figure}[h!]
\begin{center}
\includegraphics[width=9cm]{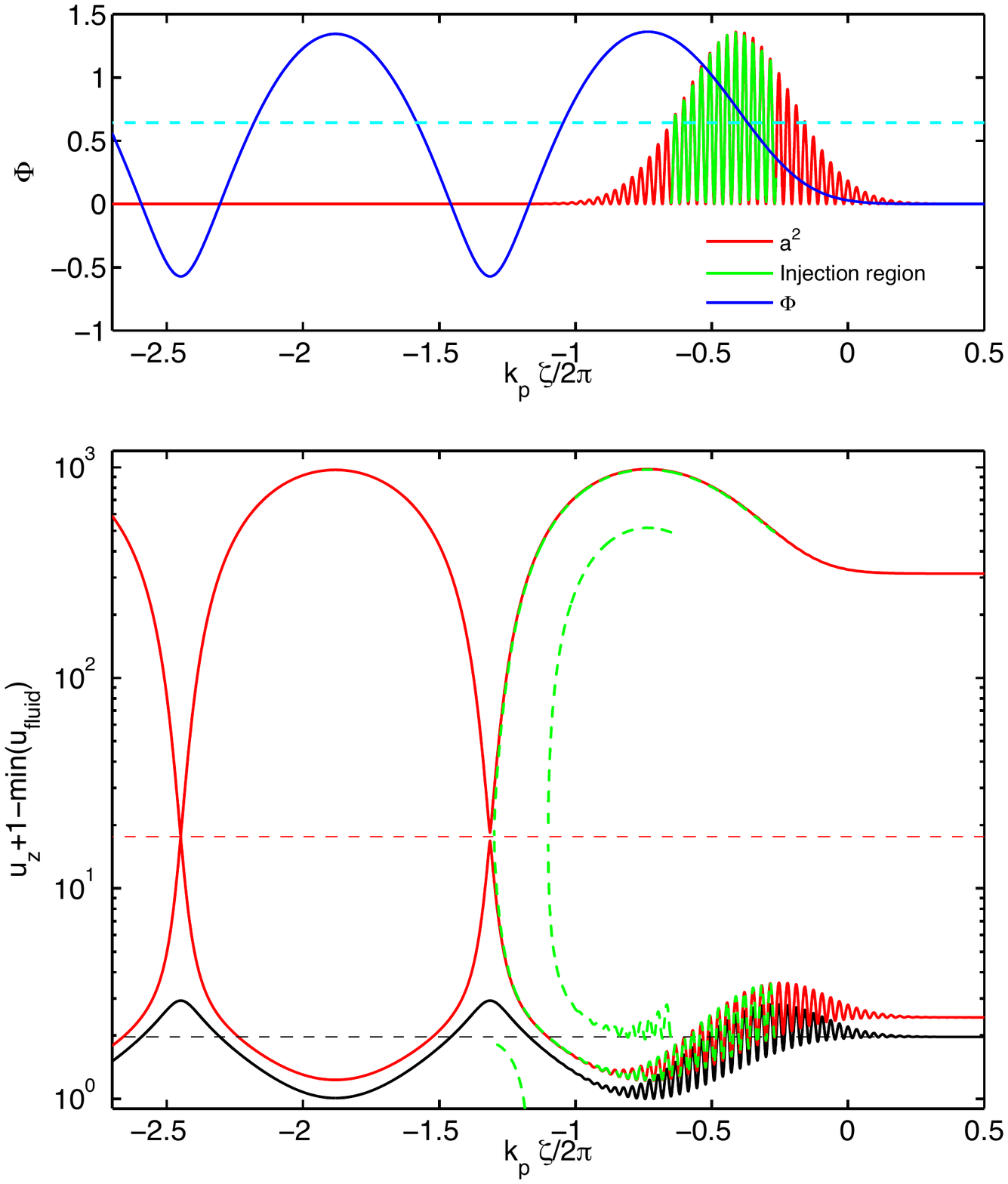}
\caption{Ionization injection---a phase space picture. The upper panel plots the laser amplitude $a^2$ (red) and the wakefield potential (blue); the dashed light-blue horizontal line indicates the ionization threshold $a_{\rm thresh}=1.37$, corresponding here to $I_{\rm thresh}=4\times 10^{18}\:\wcm$, and the region of phase space where the injection conditions are satisfied is shown in green. The lower panel shows the fluid orbit (black), the separatrix (red), and the trajectories followed by ionized electrons at the front and back of the injection region. The parameters are $a_0=2$,~$n_{\rm e}/n_{\rm c}=~0.44\%$, $\lambda=0.8\:\mathrm{\mu m}$ and $\tau=20\:\mathrm{fs}$.}
\label{figsepioniz}
\end{center}
\end{figure}

We now determine the condition(s) under which these electrons will be trapped. We first make a few assumptions concerning ionization. In our model,  ionization is assumed to occur via barrier suppression ionization \cite{ammo86}. Considering an ionization level with energy $E_{\rm i}$, the threshold intensity required to ionize the electron is given by
 \[
 I_{\rm thresh}\,[\!\wcm]=4\times 10^9\,E_{\rm i}^4\,{\rm [eV]}/Z^{*2},
 \]
where $Z^*$ is the charge of the resulting ion. Furthermore, we assume that electrons are born at rest (which is a good approximation when considering the energy that the electron will gain in the laser and plasma field). Technically, if the electron is born at phase $\zeta_{\rm ion}$, it should witness the corresponding laser amplitude $a(\zeta_{\rm ion})$. However, in linear polarization, ionization occurs mostly at the peak of the electric field, i.e.\ $a(\zeta_{\rm ion})\simeq 0$. Therefore, in the case of an ionized electron born at rest, conservation of canonical momentum reads $u_\bot(\zeta)= a(\zeta)-a(\zeta_{\rm ion})\simeq a(\zeta)$. The initial Hamiltonian for such electrons can be found from \Eref{EqHbasic} as
\[
 H_{\rm ion}=1-\phi_{\rm ion},
\]
and the ionized electron trajectory can be computed by inserting $H_{\rm ion}$ into \Eref{EqTraj}. The conditions for trapping are simply that the intensity should be high enough  for ionizing a given electron level, i.e.\ $a(\zeta_{\rm ion})>a_{\rm thresh}$, and that the electron should be born on a trapped orbit, i.e.\ $H_{\rm ion} < H_{\rm sep}$. These conditions define the region of phase space where ionized electrons are trapped and further accelerated in the wakefield. Typically, these conditions are fulfilled only in the setting of a high-intensity laser pulse ($a>1$) and a large-amplitude plasma wave. Figure~\ref{figsepioniz} illustrates this discussion: the green area in the upper panel defines the region of phase space where electrons can be injected; the corresponding trajectories are plotted in green in the lower panel (note that the fast laser frequency was considered for this case, as electrons are born at zeros of the vector potential). As discussed previously, electrons are born at rest, oscillate in the laser field, and gain energy as they get on a trapped orbit.

The injection volume in phase space depends on the wakefield amplitude and the ionization threshold, so in principle it can be controlled via the laser intensity. The injection volume is responsible for the injected charge as well as the energy spread. Obviously, large injection volumes lead to large injected charge and larger energy spreads. On the other hand, smaller energy spreads can be obtained by reducing the injection volume at the expense of charge. Experiments have demonstrated the concept of ionization injection using nitrogen and argon \cite{pak10,mcgu10}. In practice, it is quite difficult to control the injection volume experimentally: ionization injection occurs all along the propagation, as long as the laser intensity exceeds the threshold intensity; as a result, electrons are injected continuously along the propagation and the energy distributions are quite broad \cite{pak10,mcgu10}. Narrower energy spreads have been obtained by using a two-stage laser--plasma accelerator \cite{pollo11}: in the first stage, a short nitrogen gas jet is used to inject electrons while a second gas jet is used to boost the acceleration. In Ref.~\cite{golo15}, by restricting the first stage to a $500\mic$ jet, the injection was kept relatively local, resulting in energy spreads at the 10\% level.

 To summarize our discussion on ionization injection, this method is easy to implement experimentally as it requires simply the use of a high-$Z$ gas. It is a straightforward way to increase the injected charge without reaching wave-breaking amplitudes. It usually results in increased injected beam loads, but controlling the beam quality can be rather difficult, as (i)~the injection volume is directly related to the laser intensity, and (ii)~injection tends to occur continuously along the propagation.

 \section{Colliding-pulse injection}\label{secCollid}
\begin{figure}[h]
\begin{center}
\includegraphics[width=9cm]{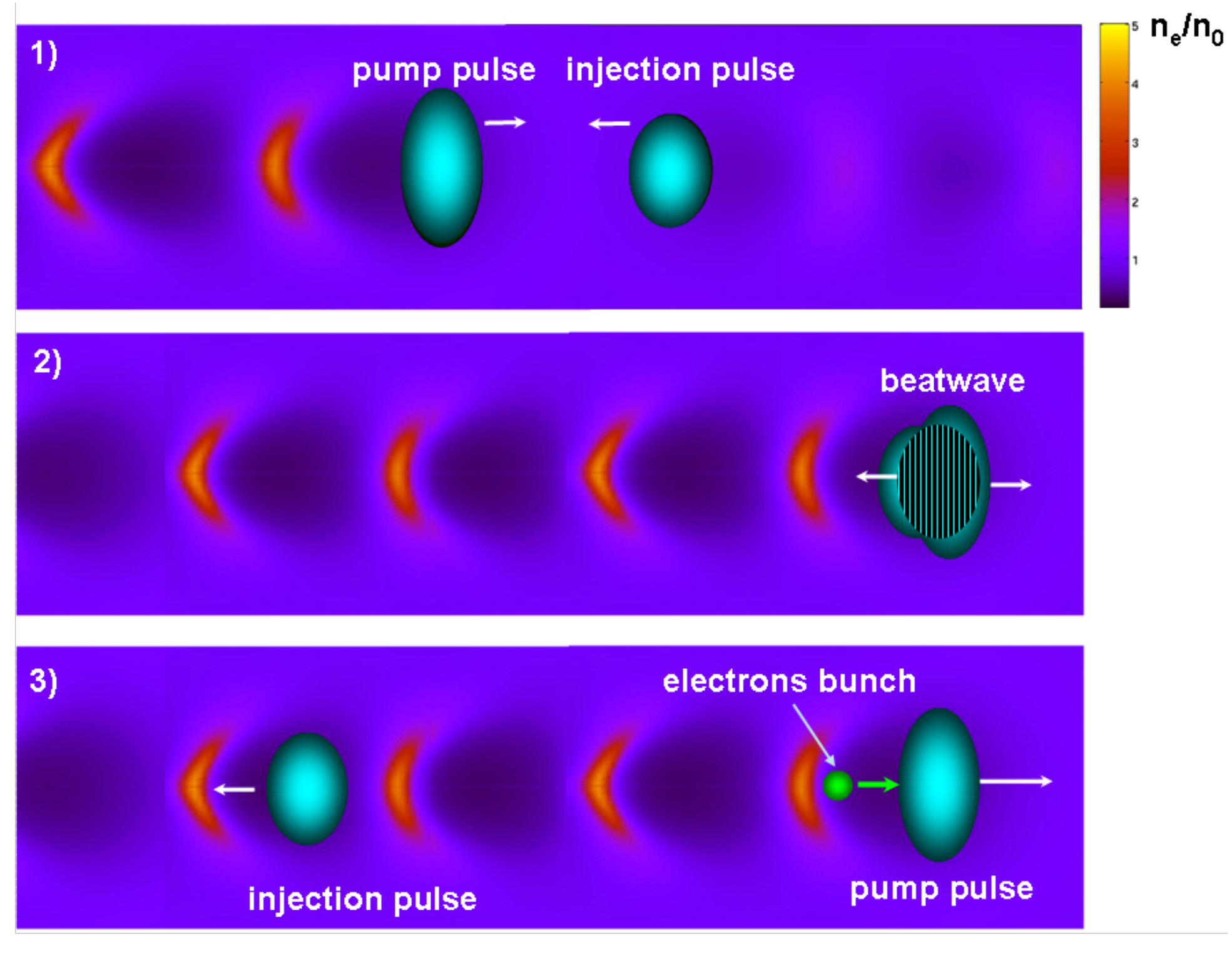}
\caption{Principle of colliding-pulse injection}
\label{figcollid}
\end{center}
\end{figure}

We now present the colliding pulse injection scheme, where an auxiliary laser pulse is used to trigger a very localized injection ($\simeq 10\mic$), resulting in beams with small energy spread. The idea of using an additional laser pulse for injecting electrons was first proposed by Umstadter \textit{et al.} \cite{umst96b}. It was further developed by Esarey \textit{et al.} \cite{esar97b}, who proposed a scheme based on counter-propagating laser pulses. In its simplest form, the scheme uses two counter-propagating ultra-short laser pulses with the same central wavelength and polarization. The first laser pulse, the `pump' pulse, creates a wakefield, whereas the second laser, the `injection' pulse, is used only for injecting electrons. The laser pulses collide in the plasma, and their interference creates a laser beat-wave pattern which can pre-accelerate plasma background electrons. If the laser intensities are high enough, this pre-acceleration permits the injection and trapping of electrons in the wakefield and their further acceleration to relativistic energies. The principle of the method is depicted in \Fref{figcollid}. Analytical work \cite{fubi04} and simulations \cite{fubi04,kota04} have shown that this two-stage acceleration mechanism can lead to the production of high-quality electron bunches with narrow energy spread, small divergence and ultra-short duration, even when using relatively modest lasers (e.g.\ with normalized vector potential $a_0=1$ for the pump pulse and $a_1=0.3$ for the injection pulse).

A theoretical description of electron injection by colliding laser
pulses was first given by Esarey \textit{et al.} in Ref.~\cite{esar97b}.
We will briefly review the principles of their
fluid model. First, the pump laser, with normalized vector potential
$a_0$, excites a plasma wave. Before the arrival of the laser pulse, the plasma
background electrons are initially at rest. In the absence of
wave-breaking and self-trapping, these electrons are not trapped in
the plasma wave. As we have seen earlier, they simply oscillate back and forth in the
plasma wave potential along the fluid orbit, and their momenta
oscillate between $u_{z,\min}^{\rm fluid}$ and $u_{z,\max}^{\rm fluid}$, where
\begin{equation}\label{eq-fluid}
    u_{z,\max,\min}^{\rm fluid}=\beta_{\rm p}\gamma_{\rm p}^2(1+\phi_{\min,\max})-\gamma_{\rm p}\sqrt{\gamma_{\rm p}^2(1+\phi_{\min,\max})^2-1},
\end{equation}
with $\phi_{\min}$ and $\phi_{\max}$ representing the minimum and maximum of the
potential, respectively.
In order to be accelerated in the
wakefield, electrons need to move along trapped orbits, inside the separatrix.
When electrons move along the
separatrix, their momenta oscillate between $u_{z,\min}^{\rm sep}$ and
$u_{z,\max}^{\rm sep}$, where
\begin{equation}\label{eq-fluid-sep}
    u_{z,\max,\min}^{\rm sep}=\beta_{\rm p}\gamma_{\rm p}(1+\gamma_{\rm p}\Delta\phi)\pm\gamma_{\rm p}\sqrt{(1+\gamma_{\rm p}\Delta\phi)^2-1}
\end{equation}
with $\Delta\phi=\phi_{\max}-\phi_{\min}$.

Therefore, one needs to find a way to push plasma background
electrons, which follow the fluid trajectories, into trapped
trajectories. The interference of two laser pulses generates a
beat-wave which is able to heat electrons and provide just such a
mechanism. The laser pulses are represented by
their normalized vector potentials $\mathbf{a}_{0,1}$, where
the subscripts 0 and 1 represent, respectively, the pump pulse and the
injection pulse. Assuming that the lasers are counter-propagating
along the $z$ direction, one can write the normalized vector
potential as
\begin{equation}\label{eq-defa}
\mathbf{a}_{0,1}=\frac{a_{0,1}(\zeta_{0,1})}{\sqrt{1+\sigma}}\bigl[\cos(k_{0,1}\zeta_{0,1})\mathbf{e}_x+\sigma\sin(k_{0,1}\zeta_{0,1})\mathbf{e}_y\bigr],
\end{equation}
where $\sigma=0$ for linear polarization and $\sigma=1$ for
circular polarization, $k_0=-k_1$ are the wavevectors, and
$\zeta_{0,1}=z\pm v_{\rm g}t$. Although in general the two waves can have different frequencies, all experiments to date have used identical frequencies, so here we assume that the two laser pulses have the same frequency~$\omega_0$.

\begin{figure}[h]
\begin{center}
\includegraphics[width=9cm]{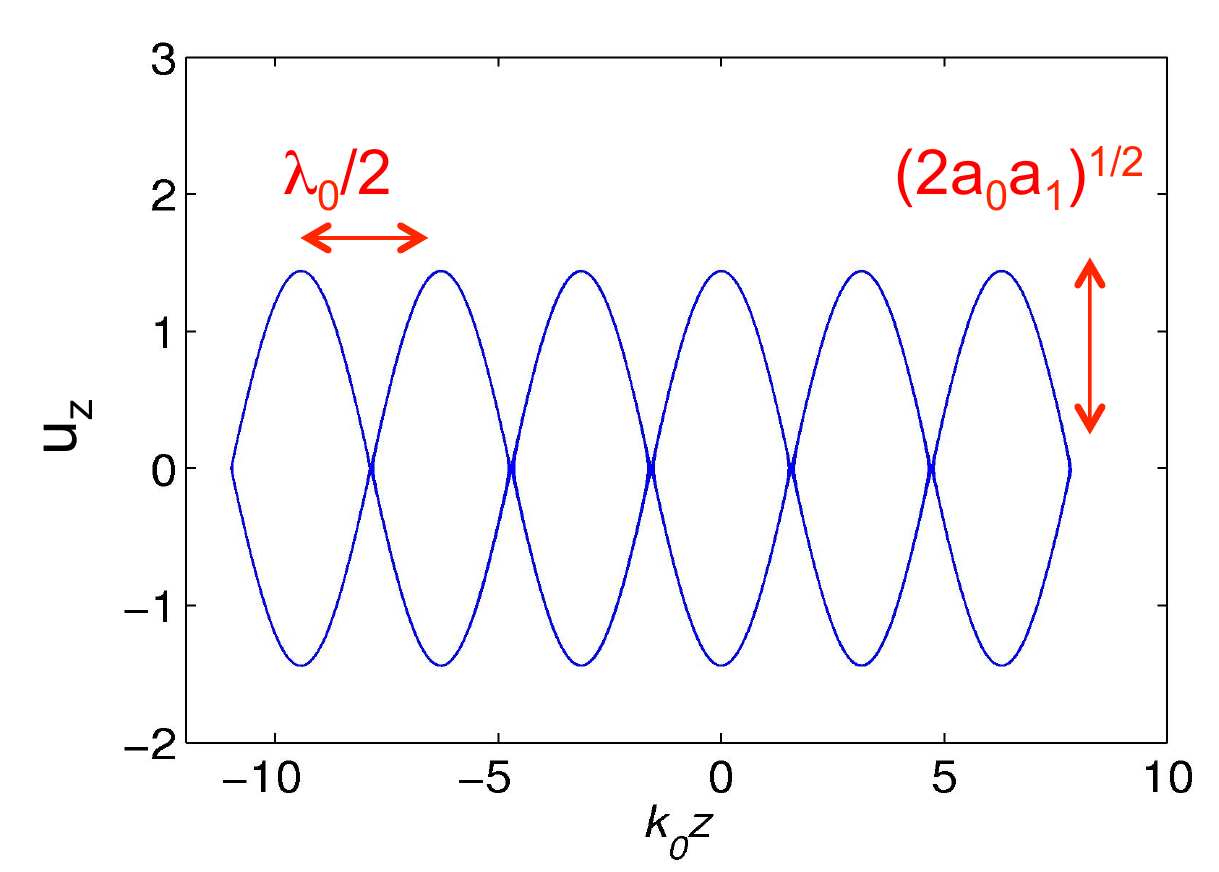}
\caption{Trajectories of electrons following the beat-wave separatrix at the collision of the two circularly polarized laser pulses.}
\label{figbeatsep}
\end{center}
\end{figure}

When the two pulses overlap, they interfere, and the resulting
squared electromagnetic field can be written as
$\mathbf{a}^2=(\mathbf{a}_0^2+\mathbf{a}_1^2+2\mathbf{a}_0\cdot \mathbf{a}_1)/(1+\sigma)$. The last term is
the beat-wave term, and it cancels out for crossed polarizations. In
the case where the two polarizations are parallel, it generates a
standing wave (i.e.\ the beat-wave has zero phase velocity) with a
spatial scale of $\lambda_0/2$. The ponderomotive force in
the beat-wave is very large (because $F_{\rm beat}\simeq 2 k_0 a_0 a_1$), and
it can pre-accelerate the plasma electrons. Neglecting the plasma
potential, electron trajectories in the beat-wave are governed by
the Hamiltonian
\begin{equation}\label{eq-Hb}
    H_{\rm b}=\sqrt{1+u_z^2+\mathbf{a}^2}.
\end{equation}
In order to obtain an analytical estimate, we will assume circular
polarization ($\sigma=1$) for the pump and injection beams, so that
$\mathbf{a}^2=(a_0^2+a_1^2)/2+a_0a_1\cos(2k_0z)$. The separatrix in
this beat-wave pattern is then given by
\begin{equation}\label{eq-sepb}
    u_{z}^{\rm beat}=\pm\sqrt{a_0a_1(1-\cos{2k_0z})};
\end{equation}
so on the beat-wave separatrix, electrons oscillate between
$u_{z,\min}^{\rm beat}$ and $u_{z,\max}^{\rm beat}$ (see \Fref{figbeatsep}), where
\begin{equation}\label{uzmin-b}
    u_{z,\max,\min}^{\rm beat}=\pm\sqrt{2a_0a_1}.
\end{equation}
For instance, an electron trapped in the beat-wave created by two laser pulses with $a_0=2$ and $a_1=0.3$ can gain about $200$~keV, which is sufficient for getting trapped in the wakefield.

One can define an approximate threshold for injection into the wakefield by applying a phase space separatrix overlap condition \cite{esar97b,fubi04}. Specifically, island overlap requires the following conditions:
\begin{align}
u_{z,\max}^{\rm beat} & >   u_{z,\min}^{\rm sep},\nonumber\\[-7pt]
& \label{eqcond} \\[-7pt]
u_{z,\min}^{\rm beat} & <  u_{z,\min}^{\rm fluid}.\nonumber
\end{align}

\begin{figure}[h]
\begin{center}
\includegraphics[width=14cm]{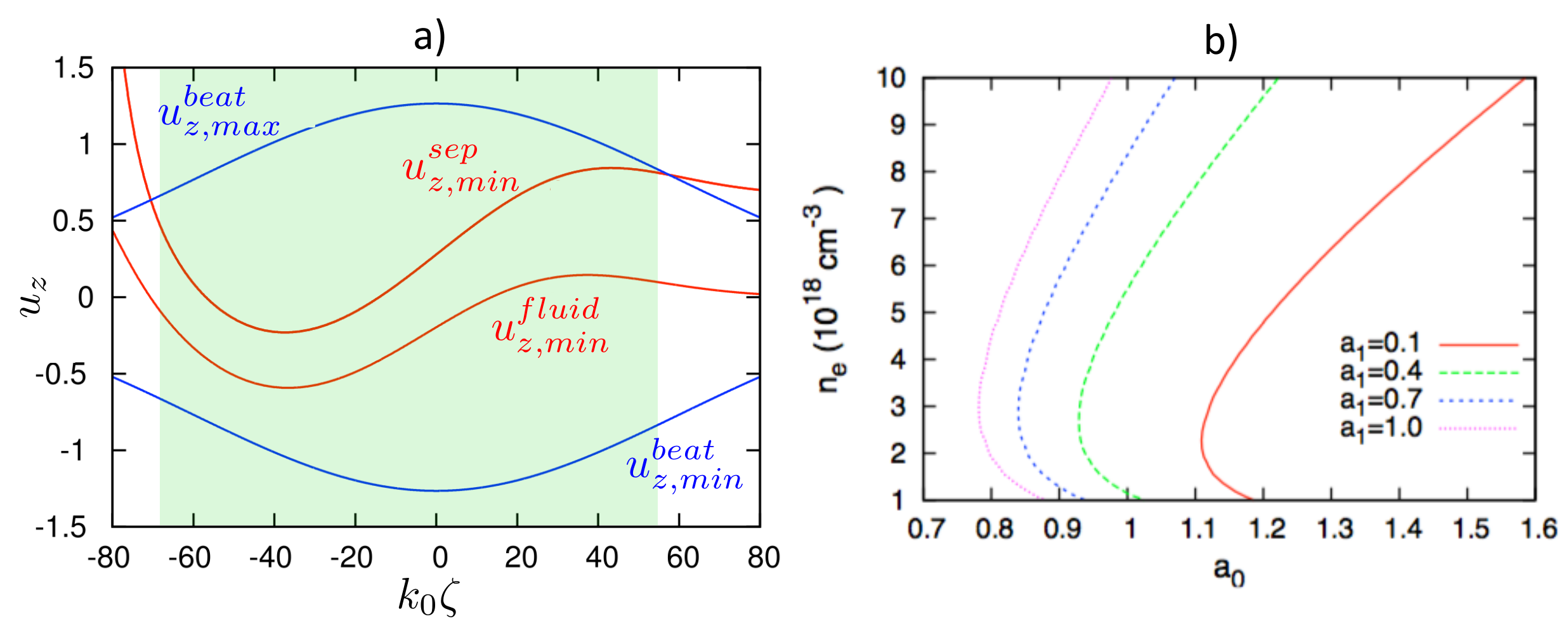}
\caption{(a) Phase space representation of the injection conditions in \eqref{eqcond} in the case of colliding pulses: the blue lines represent the amplitude of the beat-wave separatrix, while the red lines are the wakefield separatrix and the fluid orbit; the area shaded in green represents in injection volume. (b)~Plots of the injection thresholds for various laser and plasma parameters (a 30~fs and 800~nm laser pulse was assumed).}
\label{figcollidthresh}
\end{center}
\end{figure}

These conditions are illustrated in \Fref{figcollidthresh}(a). Using criterion \eqref{eqcond}, we find that the injection threshold can be reached with $a_0\simeq 1$ and $a_1\simeq 0.1$ (for this estimation, we have used pulse durations of $\tau=30$~fs to calculate the plasma wave amplitude $\phi$), for plasma density in the range of $n_{\rm e}=10^{18}$--$10^{19}\:\mathrm{cm^{-3}}$; see the results in \Fref{figcollidthresh}(b).

When the pulse polarizations are linear, the Hamiltonian in \Eref{eq-Hb} is time-dependent and no longer integrable. It has been shown that in this case, electron motion is stochastic \cite{mend83,shen02,bour05} and the simple calculations above do not hold. In fact, stochastic heating turns out to be an efficient mechanism for heating and subsequently injecting electrons \cite{shen02,davo08}. Other effects, such as plasma wave inhibition \cite{rech07}, make the physics of colliding pulses more complicated than what this simple model tends to suggest. Full 3D particle-in-cell simulations are necessary to fully capture all the physical ingredients \cite{davo08}, but the concepts presented here still hold.

The colliding-pulse injection method requires some heavy experimental investment, as a second intense laser pulse is needed. The two laser pulses need to be precisely synchronized at the femtosecond level and spatially overlapped at the micron level. Despite its complexity, colliding-pulse injection has intrinsic advantages, some of which are summarized as follows.
\begin{itemize}
\item The injection is very localized as it occurs only during the collision of the two laser pulses, i.e.\ $10\mic$ for 30~fs pulses. This helps to reduce the energy spread and to inject very short electron bunches. The injected electron beams were measured to be only a few femtoseconds in duration~\cite{lund11}.
\item The physics of injection is largely determined by the heating at the collision, so injection is less sensitive to intensity fluctuations of the main laser pulse (though the intensity of the main pulse still plays a role, as it determines the wakefield amplitude). First experiments have shown that stable beams with narrow energy spreads can be achieved in this way \cite{faur06,faur07}.
\item The location of the collision and hence the injection can be easily controlled. Therefore it is possible to tune the acceleration length. Experiments have shown that by controlling this parameter, the beam energy can be controlled over a wide range, typically 10--200~MeV \cite{faur06}.
\item The injection volume can be controlled by tuning the intensity of the injection laser pulse. For high $a_1$, the electron heating at the collision is large and results in large injected beam loads as well as larger energy spreads. The energy spread can be tuned and reduced at the 1\% level by decreasing the injection laser amplitude, as demonstrated experimentally in Ref.~\cite{rech09b}.
\end{itemize}

Finally, based on theoretical studies, a three-pulse scheme has been proposed where the main beam is  used only to generate the wakefield \cite{esar97b}; two auxiliary beams are used for injection only. Their polarization is orthogonal to the main beam polarization so that the two injection pulses do not interfere with the main beam. The advantage of this method is that it provides an additional tuning knob: the injection phase can now be chosen by tuning the distance between the main pulse and the collision point. More recently, several variations of the colliding-pulse scheme have been proposed; the basic concepts remain similar but the analysis extends to the 3D case \cite{davo09,lehe13}.

\section{Injection in density gradients}\label{secDensity}
We now focus on another scheme, in which the plasma needs to be engineered: by tailoring the plasma density, one can gain some control over the plasma wave phase velocity. As we have seen earlier, the lower the phase velocity, the lower the trapping threshold. Therefore, by setting up a local decrease in the phase velocity, one can trigger injection in a local manner as well \cite{bula98,suk01,bran08}. This can be achieved by sending the laser pulse through a downward density ramp, which causes the wakefield to slow down. A full Hamiltonian description of this problem is beyond the scope of this article; in a density transition, the wakefield potential also depends on $z$, as $\phi(z,\zeta)$, and the constants of motion that we have derived in Section~\ref{secHamiltonian} no longer hold in this case. Consequently, we will develop a simple fluid model that will provide some physical intuition of this process.

We start with the fluid equation describing the excitation of the wakefield by an intense laser pulse, restricting ourselves to the weakly relativistic case where $a^2\ll1$. We consider a gentle density gradient, $\frac{1}{n_0}\frac{\partial n_0}{\partial z}\ll k_{\rm p}$, or $L_{\rm g}k_{\rm p}\ll1$, where $L_{\rm g}$ is the gradient scale length. In this case, the plasma wave equation reads
\begin{equation}
\left(\frac{\partial^2}{\partial t^2}+\omega_{\rm p}^2(z)
\right)\phi=\omega_{\rm p}^2(z)\frac{\langle a^2\rangle}{2}.
\end{equation}
Note the spatial dependence of the plasma frequency in the gradient, $\omega_{\rm p}(z)\propto \sqrt{n_0(z)}$. We can now perform the usual change of variables to solve the problem in the moving frame: $\zeta=z-v_{\rm g}t$ and $\tau=t$. By applying the quasi-static approximation, one can then neglect the $\partial /\partial\tau$ terms. The previous equation, written in the co-moving variables, becomes
\begin{equation}
\left(\frac{\partial^2}{\partial \zeta^2}+k_{\rm p}^2(z)
\right)\phi=k_{\rm p}^2(z)\frac{\langle a^2\rangle}{2}
\end{equation}
where $k_{\rm p}(z)=\omega_{\rm p}(z)$. This equation can be integrated, and the solution behind the laser pulse has the form
\begin{equation}\label{eqphigrad}
\phi(\zeta,z)=\phi_0(z)\sin[k_{\rm p}(z)(z-v_{\rm g}t)],
\end{equation}
where the wakefield amplitude is $\phi_0(z)=-\frac{\sqrt{\pi}}{4}a_0^2(z)k_{\rm p}(z)L_0\exp\{-k_{\rm p}(z)^2L_0^2/4\}$ and its phase is $\varphi=k_{\rm p}(z)(z-v_{\rm g}t)$, so that one can compute the local oscillation frequency and wavevector:
\begin{align}
\omega & = -\partial \varphi/\partial t=k_{\rm p}(z)/v_{\rm g}=\omega_{\rm p}(z),\\
k & = \partial \varphi/\partial z=k_{\rm p}(z)+\partial k_{\rm p}/\partial z (z-v_{\rm g} t).
\end{align}
It follows that in a downward density gradient ($\partial k_{\rm p}/\partial z<0$ and $z-v_{\rm g}t <0$ behind the laser pulse), the wavevector increases with time (alternatively, the plasma wavelength decreases with time). In contrast, the plasma frequency does not depend on time: $\omega=\omega_{\rm p}(z)$. As a result of this time-varying wavevector, the phase velocity $v_{\rm p}(z,t)=\omega_{\rm p}(z)/k(z,t)$ is
\begin{equation}\label{eqvp}
v_{\rm p}(z,t) = \frac{v_{\rm g}}{1+(z-v_{\rm g}t)\dfrac{1}{k_{\rm p}}\dfrac{\rmd k_{\rm p}}{\rmd z}}.
\end{equation}
Consequently, as the wavevector increases with time, the phase velocity decreases. Injection occurs behind the laser pulse when the wakefield becomes slow enough to trap plasma background electrons.

\begin{figure}[h]
\begin{center}
\includegraphics[width=16cm]{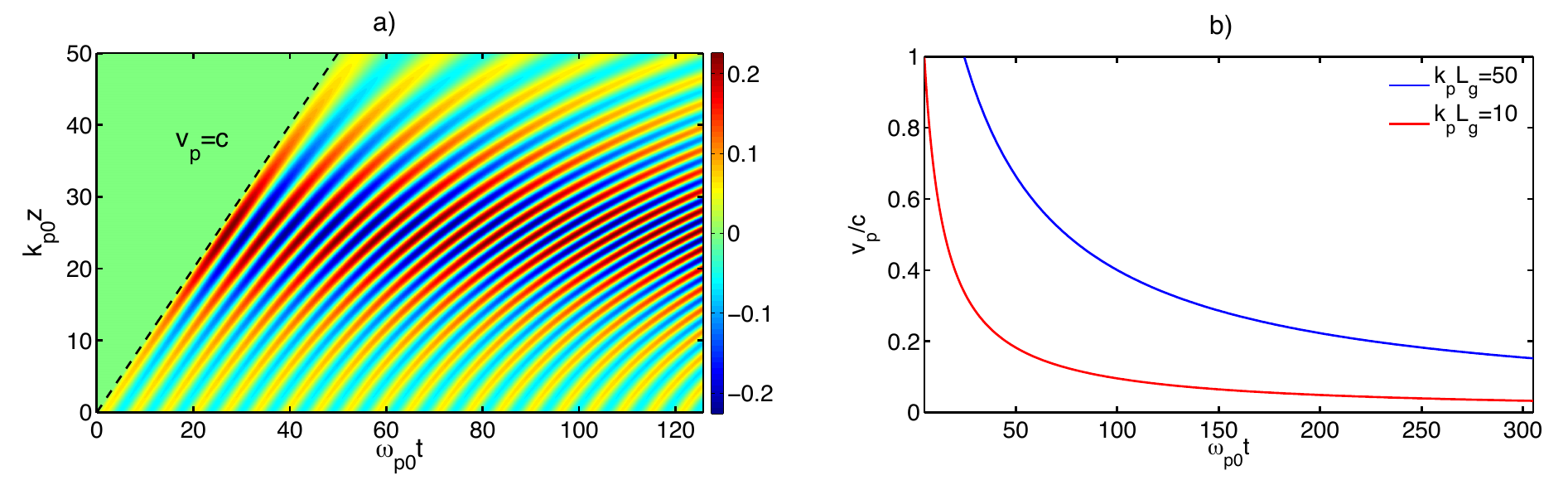}
\caption{(a) Wakefield potential map $\phi(z,t)$ in the case where a laser pulse is focused in a density gradient; \Eref{eqphigrad} was used to compute the potential; the dashed line represents the speed of light and one can see clearly the bending of the wakefield in the $(t,z)$ plane, indicating that the phase velocity is decreasing with time. (b)~Illustration of the decrease in the phase velocity with time for two different gradient scale lengths $L_{\rm g}$ (computed using \Eref{eqvp}).}
\label{figgrad}
\end{center}
\end{figure}

These effects are illustrated in \Fref{figgrad}. In panel (a), we display the wakefield potential in the case where a laser pulse is tightly focused into a density gradient. The density gradient has a Gaussian shape and $k_{\rm p0}L_{\rm g}=50$, where $k_{\rm p0}=\omega_{\rm p0}/v_{\rm g0}$ is defined using the maximum density at the top of the downward ramp. One can see clearly that the plasma frequency decreases with $z$ as the density decreases, as expected. In addition, while the initial phase velocity is close to $c$, the bending of the wakefield in the $(t,z)$ plane indicates that the phase velocity decreases with time. In panel~b), the phase velocity is plotted for various density scale lengths $L_{\rm g}$. It can be seen that the phase velocity decreases faster for sharper gradients, indicating that trapping is likely to occur more rapidly in this case.

Numerous experiments have demonstrated that trapping in density gradients is an efficient method for injecting electrons into wakefields \cite{gedd08,faur10,schm10,he13}. This controlled injection scheme has resulted in more stable beams \cite{gedd08,schm10} with energy spreads in the $10\%$ range. Some experiments have utilized the density down ramp at the exits of short gas jets, resulting in the production of low-energy beams of a few hundred keV \cite{gedd08,he13}. Other research groups have engineered a sharp density gradient by creating a shock in the gas flow \cite{schm10,buck13} or by using another laser pulse to create a density perturbation \cite{faur10}. In this case, the injection location can be controlled and the resulting beam energy can be tuned \cite{buck13,brij12}.

\section{Conclusion}
In conclusion, plasma injection schemes are at the forefront of current research in laser--plasma accelerators. Researchers continue to propose new injection schemes to test new ideas and to provide high-quality beams while maintaining a relatively simple experimental set-up \cite{bour13,yu14}. The idea of combining several methods might prove useful in the future; for example, performing colliding-pulse injection in a density gradient could increase the trapped charge while still preserving the beam quality\cite{fubi06}. Similarly, combining ionization injection and density gradient injection has the potential to yield interesting results.

\section*{Acknowledgements}
J. Faure's current research on laser--plasma acceleration is supported by the European Research Council under contract no.~306708, ERC Starting Grant FEMTOELEC.


\end{document}